\documentclass[12pt]{iopart}
\usepackage{iopams}  
\usepackage{epstopdf}
\usepackage{setstack}

\usepackage{graphicx}% Include figure files
\usepackage{bm}% bold math

\begin{document}

\title{Tailored two-photon correlation and fair-sampling: a cautionary tale}
\author{J Romero$^{1}$, D Giovannini$^{1}$, D S Tasca$^{1}$, S M Barnett$^{2}$ and M J  Padgett$^1$ }
\address{$^1$School of Physics and Astronomy, SUPA, University of Glasgow, Glasgow G12 8QQ UK} 
\address{$^2$Deparment of Physics, SUPA, University of Strathclyde, Glasgow G4 ONG UK}

\ead{jacq.romero@gmail.com}

\begin{abstract} 

We demonstrate an experimental test of the Clauser-Horne-Shimony-Holt (CHSH) Bell inequality which seemingly exhibits correlations beyond the limits imposed by quantum mechanics.   Inspired by the idea of Fourier synthesis, we design analysers that measure specific superpositions of orbital angular momentum (OAM) states, such that when one analyser is rotated with respect to the other, the resulting coincidence curves are similar to a square-wave.  Calculating the CHSH Bell parameter, $S$,  from these curves result to values beyond the Tsirelson bound of $S_{QM}=2\sqrt{2}$.  We obtain $S=3.99\pm0.02$, implying almost perfect nonlocal Popescu-Rohrlich correlations. The ``super-quantum" values of $S$ is only possible in our experiment because our experiment, subtly, does not comply with fair-sampling. The way our Bell test fails fair-sampling is not immediately obvious and requires knowledge of the states being measured. Our experiment highlights the caution needed in Bell-type experiments based on measurements within high-dimensional state spaces such as that of OAM, especially in the advent of device-independent quantum protocols. 

\end{abstract}
%\pacs{}

\maketitle

The formulation of the Bell inequality in 1964 is a hallmark in physics because it highlights the incompatibility between quantum mechanics and local hidden variable (LHV) theories in a way that is accessible to experimental test \cite{Bell1964Physics}.  A plethora of experiments showing violations of Bell inequalities confirms this incompatibility \cite{Freedman1972, A.Aspect1981, Rowe2001,  Genovese2005}.  Most of these experiments were performed with photons, where perfectly efficient detection is a challenge. To interpret violations of the Bell inequality as a demonstration of nonlocality, one has to assume some form of fair-sampling, i.e., the photons detected constitute a fair sample of all the photon pairs produced by the source \cite{Genovese2005, Berry2010}. Failure to comply with this assumption opens a detection loophole.  From a fundamental physics viewpoint, it is important for this to be closed because one can design perverse LHV theories that exploit the detection loophole to violate Bell inequalities \cite{Pearle1970}. The advent of device-independent quantum protocols, where conclusions are drawn exclusively from measurement statistics, makes compliance with fair-sampling highly relevant for practical applications \cite{Acin2007device}.  One option to ensure fair-sampling  is to use very efficient detectors which surpass the threshold efficiency values needed for detection loophole-free demonstrations \cite{Garg1987, Eberhard1993}. However, there are cases when even perfectly efficient single-photon detectors do not guarantee fair-sampling. A subtle choice of measurement states in order to obtain a desired correlation curve, as in our case, can lead to explicit violation of the fair-sampling assumption. Moreover, our choice of measurement states in this work leads to  tuneable, seemingly ``super-quantum" violations above the Tsirelson bound \cite{Cirelson1980, Chefles1996}, which is interesting from the perspective of nonlocal boxes and communication complexity \cite{Popescu1994, Brassard2006}.

\begin{figure}[h!]
\begin{center}
\includegraphics{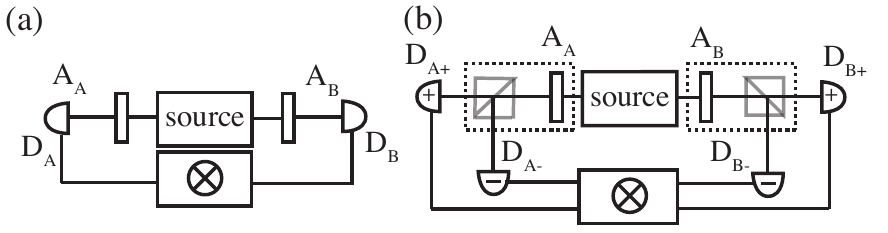}
\caption{(a) Two-channel Bell experiment as originally proposed. (b) Four-channel Bell experiment of Aspect et al. \cite{A.Aspect1981}.}
\label{fig_schematic}
\end{center}
\end{figure} 

In this work, we limit ourselves to the Clauser-Horne-Shimony-Holt (CHSH) inequality and variants thereof, which is the most tested version of the Bell inequality \cite{J.F.Clauser1969}.  The scenario considered is shown in fig.\ref{fig_schematic}.a, consisting of a source of particle pairs and two spatially separated analysers ($A_A$ and $A_B$) and detectors ($D_A$ and $D_B$). The Bell parameter, $S$, in the CHSH inequality is derived from the correlation function, $E$, of observables $O_A$ and $O_B$ as a function of the orientations (settings) of the analysers.  $O_A$ and $O_B$ can take on values of $\pm 1$ which can correspond to detection or non-detection of the photons.  The correlation function $E$ as a function of analyser settings $x$ and $y$ on arms A and B respectively can be written as
\begin{equation}
E(x,y)=P(O_A=O_B|x,y) - P(O_A\neq O_B|x,y),
\label{corr}
\end{equation}
where $P(O_A=O_B|x,y)$ is the probability that $O_A=O_B$ given that $A_A$ is set to $x$ and $A_B$ is set to $y$, similarly for $P(O_A\neq O_B|x,y)$ \cite{Scarani2013}.  More intuitively, $E$ is the difference in the probability of getting a coincidence and the probability of not getting a coincidence, given the settings $\{x,y\}$ in fig. \ref{fig_schematic}.a. The Bell parameter, $S$, is defined as $S=|E(a,b)-E(a,b')+E(a',b)+E(a',b')|$ and Bell's theorem states that LHV theories satisfy the inequality $2\geq S$, for any set of orientations $\{a,b,a',b'\}$. The inequality is violated by entangled states in quantum mechanics, which predicts that the maximum achievable value of $S$ is $2\sqrt{2}$, the Tsirelson bound \cite{Cirelson1980}. Interestingly, the Tsirelson bound is smaller than the maximum value obtained from simple algebraic arguments (which is 4);  this difference has been attributed to complementarity and compliance with information causality \cite{Cirelson1980, Chefles1996, Zukowski2009}. Hence, experiments which exhibit values of $S$ beyond Tsirelson bound would show ``super-quantum" correlations.

Because experiments can only measure count rates, the probabilities are obtained only by normalisation. Ideally, all count rates are normalised to the pair emission rate of the source, which can be measured by Ôevent-readyÕ detectors \cite{Zukowski1993}. However, they are prohibitive to implement (requiring two cascaded entangled photon sources in the experiment of \cite{Zukowski1993}) and they were not used, in particular, in the celebrated 1980s experiments of Aspect et al.  \cite{A.Aspect1981}, nor in the vast majority of experiments thereafter. Instead, in each arm, photons were sorted into the two orthogonal polarisations and sent to two different detectors.  This required four-fold coincidence counting, as shown in fig.\ref{fig_schematic}.b, and they defined the correlation function in terms of the four count rates $R_{ij}$ where ${i,j}$  correspond to the $\pm$ channels representing the two outcomes on each arm,

\begin{equation}
E(x,y)=\frac{R_{++}+R_{--}-R_{+-}-R_{-+}}{R_{++}+R_{--}+R_{+-}+R_{-+}}.
\label{corr_aspect}
\end{equation}
%\end{widetext}

In (\ref{corr_aspect}), it is natural to associate the first two terms to the probability of coincidence and the last two terms to the probability of not getting a coincidence given settings $\{x,y\}$, in the spirit of (\ref{corr}).  In the ideal case of perfect detection, these two correlation functions are the same if the denominator of  (\ref{corr_aspect}) is the same as the emission rate of the source. Because polarisation is a degree of freedom described in a two-dimensional state space, it seems only natural to accept that this is true.  In this ideal regime,  as Clauser notes, Aspect et al.'s experiments were more similar to Bohm's \emph{gedankenexperiment}, and therefore close to the original idealised experiment considered by Bell in 1964 \cite{Clauser2002,Bell1964Physics}. 

For (\ref{corr_aspect}) and (\ref{corr}) to be considered equivalent, the detected coincidences (i.e. the four count rates), are assumed to be a fair sample of all the photons emitted by the source. Postselection is introduced: in calculating $E$, we only consider the detected outcomes. However, the total of the four count rates could be related in a nontrivial way to the emission rate of the source, hence (\ref{corr_aspect}) is not necessarily equivalent to (\ref{corr}). Results can be rescaled to show violation of the Bell inequality \cite{Clauser1978}.  In the literature, the normalisation in (\ref{corr_aspect}) is referred to as the CHSH normalisation, although as Clauser points out, this is ``adamantly not" the 1969 CHSH normalisation \cite{J.F.Clauser1969, Clauser2002}.  The difference between the two may be subtle, but nevertheless important \cite{Clauser2002,  Chen2006, Clauser1974, Dada2011bell}. 

``Super-quantum" correlations have been observed previously in experiments with photons. A common theme in all these experiments is the failure to comply with fair-sampling. In \cite{Tasca2009}, for example, some of the photons were intentionally discarded (making the denominator of (\ref{corr_aspect}) smaller that it should be).   Other schemes involve threshold detectors, amplification, and fake state generation which can recreate both the quantum and ``super-quantum" violations with classical light sources  \cite{Pomarico2011, Gerhardt2011}.  Our experiment is simpler, in that our conceptual schematic is exactly that  of fig. \ref{fig_schematic}.a.  There was no tampering with the photon detectors nor with the source. We just employed analysers, which instead of measuring polarisation, measure specific spatial modes.  Moreover, the ``super-quantum" values persist even for $100\%$ efficient detectors.

\begin{figure}
\begin{center}
\includegraphics{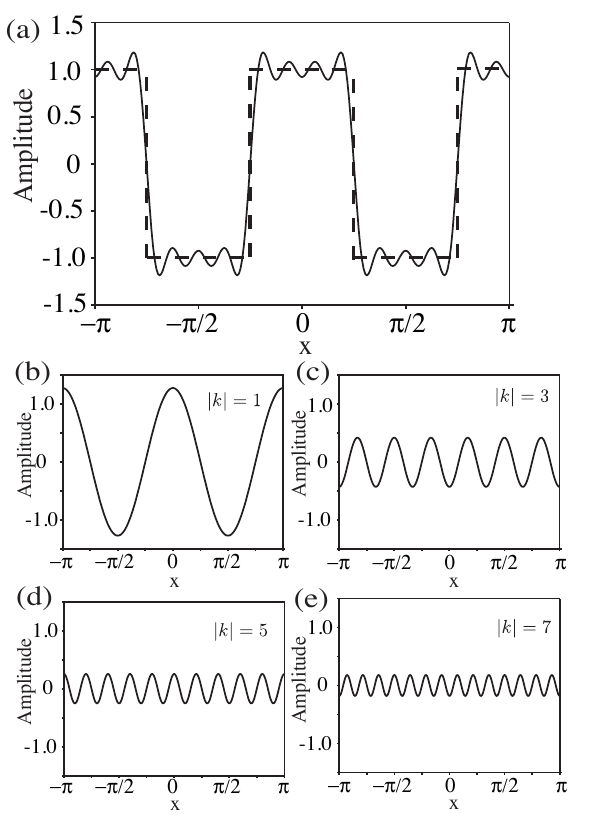}
\caption{(a) A square wave (dashed line) can be synthesised (solid line) from four appropriately weighted sinusoidal curves (b-e) corresponding to $k$-values $\{\pm1,\pm3,\pm5,\pm7\}$.}
\label{synthesis}
\end{center}
\end{figure}

Transverse spatial modes of single photons associated with optical orbital angular momentum (OAM) have steadily risen as a convenient degree of freedom for observing quantum correlations.  
We denote a photon with OAM $\ell\hbar$ as the state $|\ell\rangle$, conveniently described by the Laguerre-Gaussian family of modes, $LG^\ell_p$.  The index $\ell$ is the azimuthal index, and can take on any integer value, hence the OAM state space is, in principle, unbounded. The index $p$ is the radial index, which we set to $0$. We consider photon pairs generated via spontaneous parametric down-conversion (SPDC). As OAM is conserved in the process of SPDC, the generated two-photon state, $|\Psi_{2P}\rangle$, in the OAM basis is \cite{A.Mair2001},

\begin{equation} 
|\Psi_{2P}\rangle=\sum_{\ell=-\infty}^{\ell=+\infty}c_{\ell}|\ell\rangle_A|-\ell\rangle_B.
\label{2P}
\end{equation}
The quantity $|c_\ell|^2$ is the probability that the photon in arm A is in the state $|\ell\rangle_A$ and the photon in arm B is in $|-\ell\rangle_B$.  
Bell tests on entangled photons, exploiting OAM, have been implemented via the measurement of  equal-amplitude, coherent superpositions of opposite-valued OAM modes exhibiting $|\ell|-$ fold symmetry described by the state $|\alpha_\ell(\theta)\rangle$,
\begin{equation}
|\alpha_\ell(\theta)\rangle=(e^{i\ell\theta}|\ell\rangle + e^{-i\ell\theta}|-\ell\rangle)/\sqrt{2}. 
\label{alpha}
\end{equation}
We refer to$|\alpha_\ell(\theta)\rangle$ as 2$|\ell|$-sector states, oriented at an angle $\theta$, and characterised by $2|\ell|$ sectors of alternating $0$ and $\pi$ phase sectors. If sector states are measured in a Bell-type experiment, the resulting coincidence curve is a sinusoidal function of the relative orientations of the sector state analysers, with a period of $2\pi/|\ell|$ \cite{J.Leach2009}. The direct relationship between $\ell$ and the period of the coincidence curves is convenient in the light of Fourier synthesis, wherein an arbitrary signal can be synthesised by an appropriate weighting of sinusoidal curves of varying frequencies. An important insight in this work is the fact that we can consider the sinusoidal correlations between sector states of different $\ell$-values as a library with which to create any arbitrary two-photon correlation (fig. \ref{synthesis}). 
This concept applies to any  coincidence curve, but here we focus on synthesising a square wave-like coincidence curve because (upon postselection) this will give the algebraic maximum value of $S$, equal to $4$, just as in the example of Popescu and Rohrlich  \cite{Popescu1994}. 

Any function $g(x)$ with period $T$ and frequency $\omega_0$, can be written in a Fourier series as \cite{Goodman2005Fourier},
\begin{equation}
g(x)=\sum_{k=-\infty}^{k=+\infty} f_k e^ {i\omega _0kx},
\label{fourier}
\end{equation}
where the coefficients $f_k$ are given by 
\begin{equation}
f_k=\frac{1}{\pi}\int_{-T/2}^{T/2}g(x)e^{-i\omega _0kx}dx.
\end{equation}
We have chosen the exponential form of the Fourier series to highlight the similarity between a general superposition of the sector states in (\ref{alpha}) and a general Fourier decomposition of any function.

\begin{figure}
\begin{center}
\includegraphics{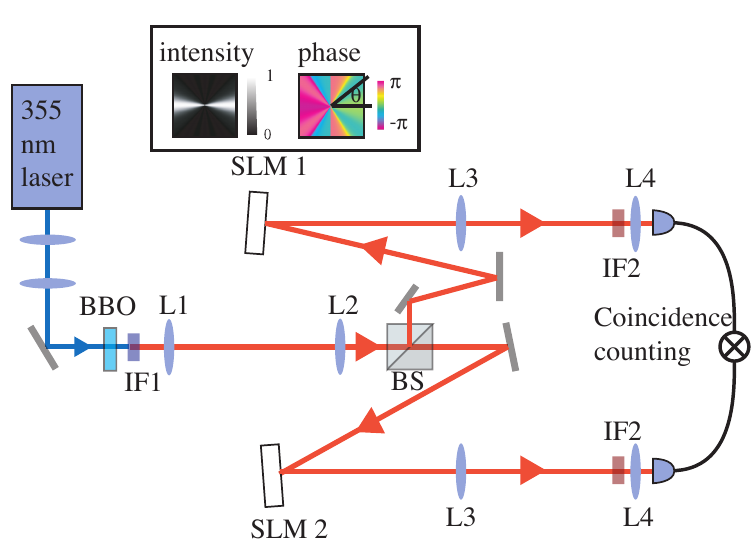}
\caption{Experiment setup. The insets show the intensity and phase of $\psi_{\theta}$, with the angle $\theta$ defining the orientation.}
\label{fig_settings}
\end{center}
\end{figure}

For the square wave in fig. \ref{synthesis} (dashed line), $\omega_0=2$, and coefficients are nonzero only for the odd values of $k$ (fig. \ref{synthesis}). Hence we consider the superposition of four odd sector states, 
\begin{equation}
|\psi(\theta)\rangle=\sum_\ell b_\ell|\alpha_\ell(\theta)\rangle,
\label{setting}
\end{equation} 
where $\{\ell\}=\{1,3,5,7\}$, $\theta$ is the azimuthal angle and $b_\ell$ are complex coefficients, $\{b_1\approx0.778, b3\approx0.467, b5\approx0.389i,b7\approx0.155\}$ which have been found by optimising experimentally. The transverse intensity and phase profiles of this mode is shown in the inset of fig. \ref{fig_settings}. A $\pi/2$ rotation of the analyser for $|\psi(\theta)\rangle$ allows us to measure $|\psi(\theta+\pi/2)\rangle$, which is orthogonal to $|\psi(\theta)\rangle$. For brevity we will refer to these as $|\psi^\bot\rangle$ and $|\psi\rangle$, respectively.

Fig.\ref{fig_settings} shows our experiment apparatus. We pump a 3 mm-thick $\beta-$barium borate (BBO) crystal cut for type-I collinear SPDC with a 355 nm fundamental Gaussian beam to produce entangled photon pairs at 710 nm.  The pump beam is blocked by an interference filter (IF). The signal and idler photons are then separated by a beamsplitter (BS), and directed to spatial light modulators (SLMs).  These are programmable devices which allow us to measure any arbitrary superposition of OAM modes, with the appropriate phase and amplitude modulation \cite{Davis1999Encoding}. The state in (\ref{setting}) is encoded on the SLMs for both arms. The SLMs are in the image plane (L1 and L2) of the exit face of the crystal, and the diffractive optics they display serve as our analysers.  The SLMs are also in the image plane of the facets of single-mode fibres (L3 and L4), which are connected to avalanche photodiode detectors. The output of the detectors are fed to a coincidence counting circuit, and we monitor the coincidences as a function of the modes displayed on the SLM.  We have subtracted the accidental coincidences, $A=C_s\times C_I\times \Delta t$, where $C_S$ and $C_I$ are the single channel counts of the signal and idler arms respectively and $\Delta t$ is the coincidence timing window (typically $\approx 5\%$ of the total counts). 

\begin{figure}
\begin{center}
\includegraphics{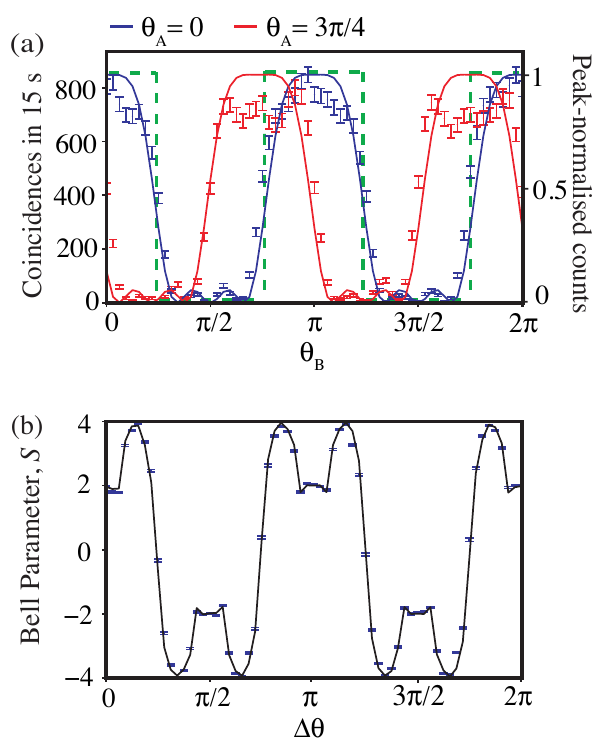}
\caption{(a) Sample coincidence curves as a function of $\theta_B$ from our experiment (blue and red dots) for two values of $\theta_A$, and fit from theory (blue and red lines).  The left scale gives the raw coincidences, and the right scale is normalised with respect to the maximum value. A square coincidence curve is also shown for reference (dashed green line). (b) We achieve $S$ above the Tsirelson bound, using the normalisation in (\ref{corr_aspect}). Our experimental values (blue dots) agrees very well with theory (black line), which predicts almost perfect Popescu-Rohrlich correlations with postselection.}
\label{fig_results}
\end{center}
\end{figure}

The coincidences as a function of the orientations of the analyser in arm B ($\theta_B$) for two orientations of the analyser in arm A ($\theta_A=0, 3\pi/4$), are shown in fig.\ref{fig_results}.a.  Allowing normalisation with respect to the maximum values, we fit our experimental coincidence curve with the expected coincidences as a function of the relative orientation, $\Delta\theta=\theta_A-\theta_B$.  This is given by $C_{th}(\Delta\theta)=|_A\langle\psi(\theta)|_B\langle\psi(\theta+\Delta\theta)|\Psi_{2P}\rangle|^2$ (right scale in fig.\ref{fig_results}.a, solid lines), assuming an ideal OAM spectrum where all the $|\ell\rangle$ states are equally weighted.  In contrast to the usual Bell-type experiments, the coincidence curve is not sinusoidal, but instead, is similar to a square wave, as we have intended (fig.\ref{fig_results}.a, green dashed line).  We do not have the equivalent of a polarising beamsplitter which distinguishes between $|\psi\rangle$ and $|\psi^\bot\rangle$.   We only have one detector in each arm, so in order to apply the normalisation in (\ref{corr_aspect}) we have to measure $|\psi\rangle$ and  $|\psi^\bot\rangle$ sequentially, just as in \cite{J.Leach2009,Oemrawsingh2004}. Following this normalisation, we calculate the correlations and obtain the value of $S$ as a function of $\Delta\theta$ (blue dots), showing good agreement with theory (black line). We obtain $S$ values above the Tsirelson bound for a range of $\Delta\theta$, with the maximum being $S=3.99\pm0.02$, achieving almost perfect Popescu-Rohrlich correlation.

We note that the state $|\Psi_{2P}\rangle$  contains all other $\ell$ values, but because of the orthogonality of the OAM modes, the measurements we make are sensitive only to $\{\ell\}=\{\pm1,\pm3,\pm5,\pm7\}$. In practice, the OAM spectrum is not flat, the values of $|c_\ell|^2$ decrease with increasing $\ell$.  As we are dealing with low OAM values however, this fact does not skew our experiment results significantly, hence we assume an ideal OAM spectrum in calculating the theoretical curves in fig.\ref{fig_results}. 

If knowledge of the measurement states is available, we can easily check if our choice of measurement states satisfy fair-sampling, hence we can avoid this subtle manifestation of the detection loophole. However, if the experiment is in a black box and we are only presented with the coincidence curve and values for $S$, as in fig. \ref{fig_results}, we will wrongly conclude that we have violated Tsirelson's bound.  We have a two-setting, two-outcome experiment, which is what the original Bell inequality requires.  As we have mentioned, the normalisation (\ref{corr_aspect}) has an inherent assumption that the dimensionality, $D$, is 2. Placing a bound on $D$ from measurement statistics is possible in a device independent manner by using dimension witnesses \cite{Brunner2008DW}.   This requires testing Bell inequalities formulated for higher dimensions such as the Collin-Gisin-Linden-Massar-Popescu (CGLMP) inequality \cite{Collins2002}, which has been violated in \cite{Dada2011}.  Testing dimension witnesses for systems showing quantum correlations is challenging, however a witness for $D=4$ exists \cite{Cai2012}. Another way of assessing dimensionality is to calculate the inverse of the area under the peak-normalised coincidence curve \cite{Pors2008shannon, Oemrawsingh2005FOAM}.  Applying this to our coincidence curves, we obtain $D=2.21$ for our measurements.  Unlike the results of  \cite{Oemrawsingh2005FOAM} which show parabolic coincidence fringes that immediately hint at the high-dimensional nature of the measurements states, our dimensionality is not far from
 $D=2$, which is the assumption of the normalisation in (\ref{corr_aspect}). We note that $D=2.21$ for our case is not the correct dimensionality, because of the peak-normalisation involved (i.e. normalising with respect to the peak when fair-sampling is not satisfied is also problematic), but to realise the incorrectness requires one to recognise that fair-sampling is not satisfied in the first place, and this is not immediately obvious.  
Without knowledge of the measurement states, our experiment seems to be a legitimate Bell experiment, in the spirit of fig.\ref{fig_schematic}. However, if we know the measurement states, we can verify a rather perverse violation of fair-sampling.

%which will not be solved by perfectly efficient photon detectors. In Bell-type experiments, poor detector efficiency is often ignored because the losses in this case are independent of the measurement settings \cite{Berry2010}. However, this is not true in our case. 
A general CHSH-Bell experiment projects onto a state $|\beta\rangle\langle\beta|$, and onto an orthogonal state $|\beta^\bot\rangle\langle\beta^\bot|$. The part of the Hilbert space sampled by these measurements is given by the sum of these two projectors, 
\begin{equation}
\hat{H}=|\beta\rangle\langle\beta|+|\beta^\bot\rangle\langle\beta^\bot|, 
\end{equation}
and this should be independent of analyser orientation in order to satisfy fair-sampling \cite{Dada2011bell}. For a two-dimensional Hilbert space, $\hat{H}$ spans the whole state space (the sum is the identity operator, $\mathbb{I}$). We take for an example, 
$|\alpha_\ell(\theta)\rangle$, which we have defined previously and has been used in \cite{J.Leach2009} to violate the Bell inequality.   The orthogonal state to $|\alpha_\ell(\theta)\rangle$ is  $|\alpha^\bot_\ell(\theta)\rangle=|\alpha_\ell(\theta+\pi/2\ell)\rangle$. One can check that indeed, $\hat{H_\alpha}=|\alpha_\ell(\theta)\rangle\langle\alpha_\ell(\theta)|+|\alpha^\bot_\ell(\theta)\rangle\langle\alpha^\bot_\ell(\theta)|=\mathbb{I}$, which is  independent of $\theta$.

Calculating $\hat{H}_{\psi}$ for our settings, $\hat{H}_{\psi}=|\psi\rangle\langle\psi|+|\psi^\bot\rangle\langle\psi^\bot|$, as a function of a particular orientation $\theta$, results to an $8 \times 8$ matrix, which has off-diagonal elements because of the asymmetry in the OAM values (the $\ell$ values do not always sum to zero).  Just looking at the first $3\times3$ corner of the matrix, we have
\begin{equation}
\hat{H}_{\psi}=\left(\begin{array}{cccc}b_7b_7^*  &0&b_7b_3^*e^{-i4\theta}&\cdots 
 \\0  & b_5b_5^*&0&\cdots\\b_3b_7^*e^{i4\theta}&0&b_3b_3^*&\cdots \\ \vdots&\vdots&\vdots&\ddots\end{array}\right)
\end{equation}
which shows a dependence on the angle $\theta$.  As we rotate our analysers, $\hat{H}_{\psi}$ changes, and we are sampling a different portion of the Hilbert space. Consequently, because of this dependence on the orientation of our analysers, we are not satisfying fair-sampling, regardless of our light collection and detection efficiencies. Even if we used the scheme of fig. \ref{fig_schematic}.b, with a perfect sorter for the two orthogonal states $|\Psi\rangle$ and  $|\Psi^\bot\rangle$, and $100\%$ efficient detectors, we will still not satisfy fair-sampling, because of our choice of measurement states.

\begin{figure}[h!]
\begin{center}
\includegraphics{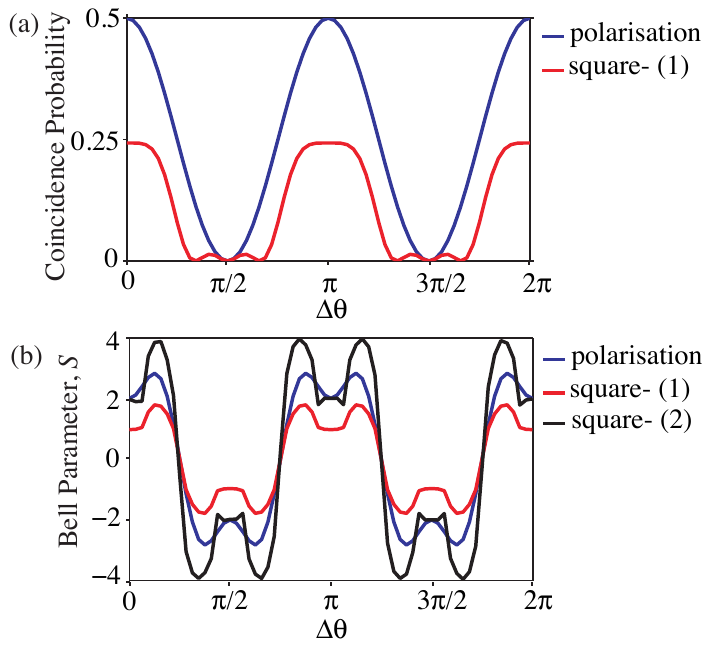}
\caption{(a) The theoretical coincidence probability for our measurement settings (red line) as a function of $\Delta\theta$ is less than that of the polarisation case (blue line). (b) Using the correct normalisation (1) , the maximum value of $S$ should have been 1.79 (red line), in contrast to $2\sqrt{2}$ for polarisation (blue line) and 3.99 for the wrong normalisation (2) (black line).}
\label{fig_correct}
\end{center}
\end{figure}

We compare the theoretical coincidence probability, of our experiment and that of a Bell-type experiment for polarisation, without renormalisation (just from the probabilities as given by Born's rule, assuming an ideal OAM spectrum) in fig. \ref{fig_correct}.a. We can obtain these curves experimentally if we know the pair emission rate of the source, but knowing this rate is very challenging. The discrepancy in the maximum probabilities between our experiment (0.24) and that of the polarisation case (0.5) points to the inherent loss that is due to our choice of measurement settings, and the fact that they are not phase-conjugates of each other, as in \cite{J.Leach2009, Romero2011}. Had the analysers been phase-conjugates of each other, the maximum probability will also be 0.5, but the shape of the coincidence curve will not be similar to a square-wave. If we follow the prescription of CHSH, and use the correlation function as defined in (1) to calculate the Bell parameter, we obtain the curves in fig.\ref{fig_correct}.b. This gives the maximum violation for our settings as $S=1.79$, in contrast to $S=3.99$, which we obtained previously by wrongly normalising our count rates.

The loss arising from the choice of measurement states is different from the loss arising from imperfect detector efficiency. The latter can be accounted for without violating fair-sampling because this loss is independent of measurement setting (orientation) \cite{Berry2010}.  This is not true for our analysers, which act in a higher-dimensional state space. A physical rotation of our analysers does not correspond to a rotation in a given plane of the Bloch sphere spanned by the states $|\psi\rangle$ and $|\psi^\bot\rangle$. This introduces an orientation-dependent postselection, which is not immediately obvious. Moreover, our tailored coincidence curves are possible only because our two-photon OAM state space, $|\Psi_{2P}\rangle$, is high-dimensional.  If $|\Psi_{2P}\rangle$ consisted of only a pair of the oppositely valued OAM states which appear in (\ref{setting})  (i.e. there is only one term in (\ref{2P}) with $\ell$ equal to 1 or 3 or 5 or 7), we will get the usual sinusoidal coincidence curves.

From the point of view of nonlocality, our experiment is a very convenient way of implementing tuneable Popescu-Rohrlich correlations  which can be viewed as a nonlocal AND gate as in \cite{Tasca2009}, with the fidelity given by $F=(S+4)/8$.  We achieve fidelities as high as $99\%$, higher than the $90.8\%$ upper bound above which communication complexity is trivial \cite{Tasca2009, Brassard2006}. We hope that this will be further explored by future studies in nonlocality and communication complexity.

In conclusion, we have shown that tailored two-photon correlation is possible with suitably designed analysers that measure specific superpositions of OAM states. 
We built on the idea of Fourier synthesis to design analysers for a Bell-type experiment, such that when one analyser is rotated with respect to the other, the resulting coincidence curves are similar to a square-wave. These coincidence curves give values of the CHSH Bell parameter which are apparently above the Tsirelson bound of $2\sqrt{2}$.  The ``super-quantum" values arise from the failure to comply with fair-sampling. The analysers that we use are not phase-conjugates of each other, and upon physical rotation,  sample different Hilbert spaces.  Hence, the way our Bell test fails fair-sampling is subtle and not immediately obvious, without knowledge of the states being measured.    Our experiment highlights the caution needed in interpreting the results of 
Bell-type experiments within high-dimensional state spaces such as that of OAM, especially in the advent of device-independent quantum protocols.

\section*{Acknowledgements}
This work is supported by EPSRC (EP/I012451/1) and Hamamatsu.  We thank J Leach, T Brougham and V Scarani for useful discussions. 

\section*{References}
\bibliographystyle{iopart-num.bst}
\bibliography{Ref}

\providecommand{\newblock}{}
\begin{thebibliography}{10}
\expandafter\ifx\csname url\endcsname\relax
  \def\url#1{{\tt #1}}\fi
\expandafter\ifx\csname urlprefix\endcsname\relax\def\urlprefix{URL }\fi
\providecommand{\eprint}[2][]{\url{#2}}

\bibitem{Bell1964Physics}
Bell J 1964 {\em Physics\/} {\bf 1} 195--200

\bibitem{Freedman1972}
Freedman S and Clauser J~F 1972 {\em Phys.~Rev.~Lett.\/} {\bf 28} 938--941

\bibitem{A.Aspect1981}
Aspect A, Grangier P and Roger G 1981 {\em Phys.~Rev.~Lett.\/} {\bf 47}
  460--463

\bibitem{Rowe2001}
Rowe M~A, Kielpinski D, Meyer V, Sackett C~A, Itano W, Wayne M, Monroe C and
  Wineland D~J 2001 {\em Nature\/} {\bf 409} 791--794

\bibitem{Genovese2005}
Genovese M 2005 {\em Phys. Rep.\/} {\bf 413} 319--396

\bibitem{Berry2010}
Berry D~W, Jeong H, Stobi{\'n}ska M and Ralph T~C 2010 {\em Phys. Rev. A\/}
  {\bf 81} 012109

\bibitem{Pearle1970}
Pearle P~M 1970 {\em Phys. Rev. D\/} {\bf 2} 1418

\bibitem{Acin2007device}
Ac{\'\i}n A, Brunner N, Gisin N, Massar S, Pironio S and Scarani V 2007 {\em
  Phys. Rev. Lett.\/} {\bf 98} 230501

\bibitem{Garg1987}
Garg A and Mermin N~D 1987 {\em Phys. Rev. D\/} {\bf 35} 3831

\bibitem{Eberhard1993}
Eberhard P 1993 {\em Phys. Rev. A\/} {\bf 47} 747--750

\bibitem{Cirelson1980}
Cirel'son B~S 1980 {\em Lett. Math. Phys.\/} {\bf 4} 93--100

\bibitem{Chefles1996}
Chefles A and Barnett S~M 1996 {\em J. Phys. A-Math. Gen.\/} {\bf 29}
  L237--L239

\bibitem{Popescu1994}
Popescu S and Rohrlich D 1994 {\em Found. Phys.\/} {\bf 24} 379--385

\bibitem{Brassard2006}
Brassard G, Buhrman H, N~Linden N, M{\'e}thot A~A, Tapp A and Unger F 2006 {\em
  Phys. Rev. Lett.\/} {\bf 96} 250401

\bibitem{J.F.Clauser1969}
Clauser J~F, M~A~Horne A~S and Holt R~A 1969 {\em Phys.~Rev.~Lett.\/} {\bf 23}
  880--884

\bibitem{Scarani2013}
Scarani V 2013 {\em arXiv preprint arXiv:1303.3081\/}

\bibitem{Zukowski2009}
Paw{\l}owski M, Paterek T, Kaszlikowski D, Scarani V, Winter A and {\.Z}ukowski
  M 2009 {\em Nature\/} {\bf 461} 1101--1104

\bibitem{Zukowski1993}
{\.Z}ukowski M, Zeilinger A, Horne M~A and Ekert A~K 1993 {\em Phys. Rev.
  Lett.\/} {\bf 71} 4287--4290

\bibitem{Clauser2002}
Clauser J~F 2002 {\em Quantum Unspeakables\/} ed Bertlmannm R~A and Zeilinger A
  (Springer)

\bibitem{Clauser1978}
Clauser J~F and Shimony A 2001 {\em Rep. Prog. Phys.\/} {\bf 41} 1881

\bibitem{Chen2006}
Chen Y~A, Yang T, Zhang A~N, Zhao Z, Cabello A and Pan J~W 2006 {\em Phys. Rev.
  Lett.\/} {\bf 97} 170408

\bibitem{Clauser1974}
Clauser J~F and Horne M~A 1974 {\em Phys. Rev. D\/} {\bf 10} 526

\bibitem{Dada2011bell}
Dada A and Andersson E 2011 {\em Int. J. Quantum Inf.\/} {\bf 9} 1807--1823

\bibitem{Tasca2009}
Tasca D~S, Walborn S~P, Toscano F and Souto~Ribeiro P~H 2009 {\em Phys. Rev.
  A\/} {\bf 80} 030101

\bibitem{Pomarico2011}
Pomarico E, Sanguinetti B, Sekatski P, Zbinden H and Gisin N 2011 {\em New J.
  of Phys.\/} {\bf 13} 063031

\bibitem{Gerhardt2011}
Gerhardt I, Liu Q, Lamas-Linares A, Skaar J, Scarani V, Makarov V and
  Kurtsiefer C 2011 {\em Phys. Rev. Lett.\/} {\bf 107} 170404

\bibitem{A.Mair2001}
Mair A, Vaziri A, Weihs G and Zeilinger A 2001 {\em Nature\/} {\bf 412}
  313--316

\bibitem{J.Leach2009}
Leach J, Jack B, Romero J, Ritsch-Marte M, Boyd R~W, Jha A~K, Barnett S~M,
  Franke-Arnold S and Padgett M~J 2009 {\em Opt.~Express\/} {\bf 10} 8287--8293

\bibitem{Goodman2005Fourier}
Goodman J~W 2005 {\em {Introduction to Fourier optics}\/} (Roberts \& Company
  Publishers)

\bibitem{Davis1999Encoding}
Davis J~A, Cottrell D~M, Campos J, Yzuel M~J and Moreno I 1999 {\em Appl.
  Opt.\/} {\bf 38} 5004--5013

\bibitem{Oemrawsingh2004}
Oemrawsingh S~S~R, Aiello A, Eliel E~R, Nienhuis G and Woerdman J~P 2004 {\em
  Phys. Rev. Lett.\/} {\bf 92} 217901

\bibitem{Brunner2008DW}
Brunner N, Pironio S, Acin A, Gisin N, M{\'e}thot A~A and Scarani V 2008 {\em
  Phys. Rev. Lett.\/} {\bf 100} 210503--210503

\bibitem{Collins2002}
Collins D, Gisin N, Linden N, Massar S and Popescu S 2002 {\em
  Phys.~Rev.~Lett.\/} {\bf 88} 40404

\bibitem{Dada2011}
Dada A, Leach J, Buller G, Padgett M~J and Andersson E 2011 {\em Nat.~Phys.\/}
  {\bf 7}

\bibitem{Cai2012}
Yu C, Scarani V and Bancal J~D 2012 {\em {The 7th Conference on Theory of
  Quantum Computation, Communication and Cryptography }\/}

\bibitem{Pors2008shannon}
Pors J~B, Oemrawsingh S~S~R, Aiello A, Exter M~P~V, Eliel E~R, van~'t Hoff€™
  G~W and Woerdman J 2008 {\em Phys.~Rev.~Lett.\/} {\bf 101} 120502

\bibitem{Oemrawsingh2005FOAM}
Oemrawsingh S~S~R, Ma X, Voigt D, Aiello A, Eliel E~R, Õt~Hooft G~W and
  Woerdman J~P 2005 {\em Phys. Rev. Lett.\/} {\bf 95} 240501

\bibitem{Romero2011}
Romero J, Leach J, Jack B, Dennis M~R, Franke-Arnold S, Barnett S~M and Padgett
  M~J 2011 {\em Phys.~Rev.~Lett.\/} {\bf 106} 100407

\end{thebibliography}

\end{document}